\documentclass[]{interact}

\usepackage{epstopdf}
\usepackage[caption=false]{subfig}
\usepackage[utf8]{inputenc}
\usepackage{cite}
\usepackage{amsmath,amssymb,amsfonts}
\usepackage{algorithmic}
\usepackage{graphicx}
\usepackage{textcomp}
\usepackage{xcolor}
\usepackage{xcolor}
\usepackage[T1]{fontenc} 
\usepackage[utf8]{inputenc}
\usepackage{amsmath}
\usepackage[cmintegrals]{newtxmath}
\usepackage{graphicx}
\usepackage{tikz}
\usepackage{tcolorbox}
\usepackage{pdflscape}
\usepackage{rotating}
\usepackage{fancyhdr} 
\usepackage{color,colortbl}
\usepackage{longtable}
\usepackage{multirow}
\usepackage{graphicx}
\usepackage{enumitem}
\usepackage{fontawesome}
\usepackage{multirow}
\usepackage{balance}
\usepackage{tablefootnote}
\usepackage{colortbl}
\usepackage{hyperref}
\usepackage[rightmargin=0.3pt,leftmargin=0.3pt]{quoting}
\usepackage[export]{adjustbox}
\usepackage{framed}
\usepackage[capposition=top]{floatrow}
\usepackage{arydshln}
\usepackage{comment}

\usepackage{multirow}
\usepackage{booktabs}
\usepackage{xcolor}
\usepackage{array}

\usepackage{framed}

\AtBeginDocument{%
  }
    
\theoremstyle{plain}

\theoremstyle{definition}

\theoremstyle{remark}

\usepackage{xcolor}

\begin{document}


\title{%
\colorbox{gray!20}{%
  \parbox{\dimexpr\textwidth-2\fboxsep}{%
    \footnotesize {Abaei, G., Shahin, M., Spichkova, M., One-Year Internship Program on Software Engineering: Students' Perceptions and Educators' Lessons Learned. Preprint copy – Accepted to appear in European Journal of Engineering Education, Taylor \& Francis, 2026.}
  }%
}\\[1em]
One-Year Internship Program on Software Engineering: Students' Perceptions and Educators' Lessons Learned 
}
\author{
\name{Golnoush Abaei\thanks{CONTACT G.~A. Author. Email: golnoush.abaei@rmit.edu.au}, Mojtaba Shahin, and Maria Spichkova}
\affil{School of Computing Technologies, RMIT University, Melbourne, Australia}}

\maketitle

\begin{abstract}
The inclusion of internship courses in Software Engineering (SE) programs is essential for closing knowledge gaps and improving graduates' readiness for the software industry. Our study focuses on year-long internships at RMIT University (Melbourne, Australia), which offers in-depth industry engagement. We analysed how the course evolved over the last 10 years to incorporate students' needs and summarised the lessons learned that can be helpful for other educators supporting internship courses. Our qualitative analysis of internship data based on 91 reports during 2023-2024 identified three challenge themes the students faced, and which courses were found by students to be particularly beneficial during their internships. On this basis, we proposed recommendations for educators and companies to help interns overcome challenges and maximise their learning experience.
\end{abstract}

\begin{keywords}
Internships, Software Engineering, Work-Integrated Learning, Student Perceptions
\end{keywords}

\section{Introduction}\label{sec:intro}

The integration of internship courses into Software Engineering (SE), Computer Science (CS), and Information Technology (IT) programs has been a topic of discussion among educators for decades \cite{ziegler1987highly,schambach1997s}, focusing on the preparedness of new graduates for the software industry. Research indicates that practical training benefits both students and companies \cite{vuoriainen2025six}. Educators emphasise the importance of real industry experiences for enhancing student employability \cite{tuzun2018computer,jaradat2017internship} and commonly agree that sharing real industry experiences with students is essential for preparing them for their careers and increasing their employability \cite{shebaro2022closing, spichkova2019industry, jackson2015employability, hol2023new}. Work-Integrated Learning (WIL) provides students with opportunities to gain industry experience while pursuing their university degrees. This approach has gained popularity over the last two decades across various disciplines due to its effectiveness in enhancing students' employability skills \cite{ferns2014work, smith2012evaluating, jackson2018developing, jackson2021embedding}. In Australia, a WIL component is mandatory for the accreditation of computing degrees \cite{ACS}.

To provide WIL experiences for students, several solutions can be implemented. One effective approach is embedding \emph{capstone projects} in the curriculum, typically conducted in the final year. These team-based projects allow educators to maintain control over learning and are structured with defined activities overseen by a course coordinator~\cite{ICSE_2018_capstone, balaban2018software, knudson2018global, bastarrica2017can,berg2025new}. Another solution is organising \emph{industrial summer schools}~\cite{tuzun2018computer}, where students gain practical skills that complement their theoretical studies through structured classes like lectures, workshops, and practical labs. 
\emph{In our work}, we focus on the third approach to providing industry experience within the SE study program: \emph{internship courses}. This approach allows students to gain a deeper understanding of real-world industrial settings, where the learning environment truly resembles an actual workplace in a company. As interns, students attend the company's office, collaborate with employees, and learn from this experience, often gaining more insights than they would in a capstone course or summer school. 

Existing studies on internship courses typically aim to justify the usefulness of having internships within university study programs, see e.g., \cite{jaradat2017internship, karunaratne2019students, ocampo2020role, anjum2020impact, kapoor2019understanding}. 
The results of such studies typically demonstrate that internships are very useful in enriching students' professional skills, which is also typically the case for WIL courses in general.
Based on all existing evidence from previous studies, we believe that the usefulness of internships (and WIL in general) can be taken as an axiom. However, questions on \textbf{how} an internship course can be structured, \textbf{how} it should be embedded into the study curriculum, and \textbf{what skill set} might help students to be more successful in their internship journey (and, even more importantly, employable after their graduation) require additional elaboration. 
Delivering an \emph{effective} internship course presents significant challenges. While the aim is to maintain an authentic workplace experience with tasks typically expected of a junior software engineer, the role of the course coordinator must be limited to preserve the authenticity of the setting. 
Therefore, an internship course serves as a “two-faced Janus” in the study curriculum, effectively facilitating the \emph{transition} from a university-based study environment to an industrial one.

\textbf{Contributions:} 
In this paper, we share our experience on organising a one-year-long (two-semester-long) internship course within the degree \emph{BP096 Bachelor of Software Engineering (Professional)}\footnote{For simplicity, it will be referred to as the Bachelor of Software Engineering hereafter.} in RMIT University (Melbourne, Australia) over the last 10 years. We also explore students' perceptions of the internship program. We aim to answer the following research questions to provide recommendations to educators and companies, which would allow interns to overcome challenges and maximise their learning experience:
\begin{itemize} [leftmargin=4ex]
   \item \textbf{\textit{RQ1. What SE program's courses do students believe most contributed to preparing them for the internship program?}}
   \item        \textbf{\textit{RQ2. What challenges do students experience during the internship program?}}
    \item \textbf{\textit{RQ3. What lessons have been learned by the course coordinator team over the last 10 years?}}
    
\end{itemize}

For RQ1 and RQ2, we conducted an empirical study analysing students' perceptions of which courses contributed to their internship preparedness and the challenges faced during their internships. Data were collected through regular reports that students submitted during their internship course, allowing the course coordinator to monitor progress and intervene when necessary. These reports also encourage students to reflect on which elective courses to take in their final year after their internship in the 3rd year. Additionally, analysing students’ perceptions helps identify beneficial skills for graduates, aligning with junior roles they may pursue post-graduation. Based on our findings, we provide recommendations for educators and the industry to better align the curriculum with students' needs.

To address RQ3, we examined the course’s evolution over the past ten years and how we navigated internship-specific challenges, drawing on our experiences as course coordinators. We also share the lessons learned to support educators in enhancing their internship programs. 
RQ1 and RQ2 aim to analyse the current/recent perceptions of students regarding their preparedness for the internship and the challenges they experienced (the data has been collected over 2023 and 2024).  In contrast to RQ1 and RQ2, we aim to provide a more holistic view of the internship course by addressing RQ3. For these reasons, we reflected not only on the last two years, but also went deeper to cover the last decade. Our focus is especially on the lessons learned related to the structure of the internship course, the learning and teaching activities within the course and the corresponding engagement with the students.

It is important to note that the terms \textit{Course Coordinator} and \textit{Program Manager} are used frequently throughout this text. The Course Coordinator is responsible for overseeing individual courses, including preparing learning materials, designing and assessing assignments, and supporting students throughout the course. The Program Manager, on the other hand, oversees the design, governance, and delivery of the degree program, ensuring alignment with university policies and accreditation requirements while leading ongoing evaluation and continuous improvement activities. The Program Manager also works closely with Course Coordinators to support effective program management and ensure a high-quality student experience. In this research study, we specifically refer to the Internship Course Coordinator and the Software Engineering Program Manager.

\textbf{Paper Organization}: Section \ref{sec:background} provides background information and summarizes related work. Section \ref{sec:method} outlines our methodology for examining both students' perceptions and educators' lessons learned. It also describes the structure of the internship program at RMIT University and discusses the study’s limitations. Section \ref{sec:findings} discusses our findings, and Section \ref{sec:recommendations} provides recommendations based on those findings. Finally, Section \ref{sec:conclusions} concludes with key takeaways and future directions for research.

\section{Background and Related Work}
\label{sec:background}

Our research focuses on a year-long (two-semester) internship for SE students. We aim to explore the specific courses and challenges these students encounter during their internships, as well as the challenges observed from the side of course coordinators (and the corresponding lessons learned), rather than merely discussing broad generalities. By doing this, we hope to shed light on the nuanced experiences that can provide educators with a more comprehensive understanding of the various factors impacting student learning. Our reflections will offer valuable insights that can inform teaching practices and curriculum development in this field. To our knowledge, no prior research has been conducted in this particular context. 

In this section, we discuss research studies conducted on the related topics. 
Some studies, such as \cite{patacsil2017exploring}, discuss the differences in perspectives between students and industry partners regarding the importance of employable skills, as well as how internship experiences influence the career decisions of female students~\cite{lapan2023no}. In the field of CS/SE, several papers have examined the impact of short-term internships integrated into the curriculum \cite{jaime2019effect, kapoor2019understanding, shin2013development,nabi2025assessing}. Other studies have addressed the challenges arising from the significant increase in the number of interns \cite{sweetser2020setting, chng2018redesigning, minnes2021cs}.

\subsection{Employability Skill Development through WIL}
{ key objective of WIL courses is the development of employability skills required for professional practice in the  industry. These skills are often developed through workplace experiences such as \emph{internships} \cite{jaradat2017internship, karunaratne2019students, ocampo2020role, anjum2020impact, kapoor2019understanding}, \emph{industry-based capstone projects} ~\cite{ICSE_2018_capstone, balaban2018software, knudson2018global, bastarrica2017can,berg2025new}, and \emph{industrial summer schools} ~\cite{tuzun2018computer} which enable students to apply theoretical knowledge in real-world contexts.

The employability skills developed through these experiences include teamwork, communication, problem-solving, project management, and professional responsibility \cite{ferns2014work, smith2012evaluating, jackson2018developing, jackson2021embedding}. A systematic review of teaching methods in Software Engineering education highlighted the importance of aligning SE courses with industry-relevant skills and practices \cite{anivcic2022teaching}. Several studies have investigated how computing programs support the development of these skills. For example, surveys of IT graduates have examined their perceptions of employability skills and the extent to which their education prepared them for industry roles \cite{kalra2021meaning}. Another study comparing students and industry partners identified differences in the perceived importance of employability skills, revealing that students often undervalue management skills that industry partners consider essential \cite{patacsil2017exploring}. Internship experiences, which represent a common form of WIL, have also been shown to influence students’ professional development. For instance, interviews with 13 female CS undergraduates examined how internship experiences affect career choices and highlighted challenges faced during internships in the United States  \cite{lapan2023no}.


\subsection{Internship Delivery Structures in WIL}
Various structures have been used to deliver WIL internships in computing programs. The selection of an appropriate structure is influenced by several factors, including cohort size, availability of industry placements, supervisory resources, and assessment and coordination requirements.

The study curriculum discussed in \cite{jaime2019effect} was similar to ours, incorporating an internship before an industrial capstone project. The authors examined the impact of internships in the CS/SE curriculum at the University of La Rioja (Spain) through two surveys: one before and one after the internships were introduced. The results showed that internships improved students' skills in technology, methodologies, and project management. Additionally, the study revealed a shift in educators' roles during capstone projects, with advisors focusing more on progress monitoring and presentation skills than on technical support and time management. 

A Master-level study at the Australian National University (ANU) \cite{sweetser2020setting} examined semester-long internships (15 hours per week) with small cohorts of 14-17 students over three semesters. This small size raises concerns about scalability, as methods suitable for 15-20 students may not apply to larger groups of 100+. With rising enrolments in CS/SE/IT programs driven by industrial demand, scalability is increasingly critical. For example, internships at Sunway University (Malaysia) \cite{chng2018redesigning} grew from 61 in 2014 to 121 in 2017.

A study at the University of California \cite{minnes2021cs} examined a large-scale internship program with 346 undergraduate CS and Engineering students. These full-time, paid internships lasted 8 to 14 weeks over the summer. The authors analysed students' reflections to identify key aspects they value in their internships. The findings may help internship coordinators understand student priorities, aiding in the improvement of the internship course and curriculum.


Only one study from fields outside of CS/SE was somewhat similar to our research \cite{griffin2019business}, which explored the perspectives of undergraduate business students on the essential skills needed for business majors. It identified in-class activities that effectively prepared them for the workforce and highlighted skills that require further emphasis in business education. While we acknowledge the relevance of this study, we cannot directly relate it to our research on SE students due to differences in field, methodology, participant demographics, and internship duration. 

\section{Methodology}
\label{sec:method}

\subsection{Institutional Context and Program Structure of the Software Engineering Degree at RMIT University} \label{sec:SEdegree}

RMIT University, established in 1887 as the Royal Melbourne Institute of Technology, is a metropolitan public institution based in Melbourne, Victoria, Australia, and is internationally recognised for its strong emphasis on applied research and industry-connected teaching in disciplines such as engineering, technology, design and business. Within the university, the School of Computing Technologies delivers programs across software engineering, computer science, cybersecurity, data science, artificial intelligence and information technology, embedding project-based learning and sustained industry engagement throughout the curriculum. The \emph{Bachelor of Software Engineering} is structured as a four-year undergraduate program comprising two years of foundational study in programming, algorithms, databases, mathematics, systems architecture and the software development life cycle, followed by a year-long industry placement and a final year focused on advanced technical and professional topics, including an industry-based capstone project. Students develop expertise in software analysis and design, testing, deployment, quality frameworks and large-scale system development, while gaining experience working within complex software environments such as operating systems, communications networks, web platforms and enterprise applications. The degree does not offer separate majors, as all students are enrolled in software engineering as their principal field; however, in the final year, students may broaden their academic profile by selecting four advanced electives or minor subjects drawn from other computing programs such as computer science, cybersecurity, data science and information technology, or from multidisciplinary areas including business or engineering. Overall, the program is designed to prepare graduates for a wide range of professional roles, including software developer, software engineer, architect, tester and systems analyst, while also supporting longer-term progression into leadership and project management positions within the software industry.

\subsection{RMIT SE Internship Program: Core structure}
\label{sec:intenship}

\textbf{Goal.} 
The internship courses at RMIT University provide students with practical industry experience related to their SE degree. These courses include a WIL component, enabling students to apply their knowledge in real-world settings. Additionally, feedback from industry professionals further enriches their learning experience. Internships typically last one year, covering two consecutive semesters, and are usually completed during the third year of the four-year-long \emph{Bachelor of Software Engineering}. This is a full-time commitment of 38 to 40 hours per week. Interns generally receive four weeks of paid leave as outlined in their contracts. If students are unable to secure an internship during the first or second semester of their third year, the Program Manager may recommend that they enrol in certain fourth-year courses that do not require the completion of an internship as a prerequisite. This allows students to extend their preparation time and strengthen their skills and resumes in preparation for internship placement opportunities in the following semester.

If the internship takes place in Australia, the company must comply with Fair Work regulations, and interns should also receive superannuation. Additionally, their salary must not be less than the minimum wage in Australia. For internships outside of Australia, compensation must meet or exceed the minimum wage for a junior software engineer in the host country.

{As stated earlier, internships cannot be interrupted unless unforeseen circumstances arise. However, students have the option to use their annual leave as specified in their contracts. Interns can begin their roles in either the first or second semester of the academic year. Their break schedule should align with their contract, which generally includes provisions for annual leave and personal leave. Students can switch employers if they find a more suitable job. However, the second job should start immediately after the termination of the first, which is not recommended by the university.

{All students will spend most of their time in the workplace, with no distinction made between interns and other employees regarding work hours and benefits. Course assignment preparation should not consume a significant amount of their time; students can complete this work after hours, on weekends, or during their annual leave. The total time allocated for assignments should not exceed 12 hours per semester. Details regarding assignments are provided at the end of this section.

Student enrolments in the Software Engineering (SE) degree have grown steadily over the past decade. As shown in Figure \ref{fig:internshipEnrolments}, SE internship placements at RMIT also have increased from approximately 12–18 students to around 60–70 students per semester, which is consistent with the trends reported in \cite{chng2018redesigning}. Across the reported semesters, the cohort consisted of approximately 83\% male and 17\% female students, indicating a predominantly male student population.

Respectively, the number of enrolments in the internship courses has been growing (see Figure \ref{fig:internshipEnrolments}; \emph{S1} and \emph{S2} denote Semester 1 and Semester 2). The primary objective of the internship course is to provide students with a chance to become familiar with the latest trends and practices of software development in the industry. Additionally, the course aims to help students understand the professional aspects associated with SE. Furthermore, students will reflect on their work placement experiences and put them in the context of their personal growth. Upon successful completion of the course, students will be equipped to work across diverse IT platforms, critically evaluate industry tools and solutions, communicate effectively in both written and oral forms, manage professional relationships, and adapt responsibly to evolving workplace conditions. They will also demonstrate a strong understanding of industry standards and processes relevant to their field.
\begin{figure*}
    \centering
    \includegraphics[width= 1\linewidth]{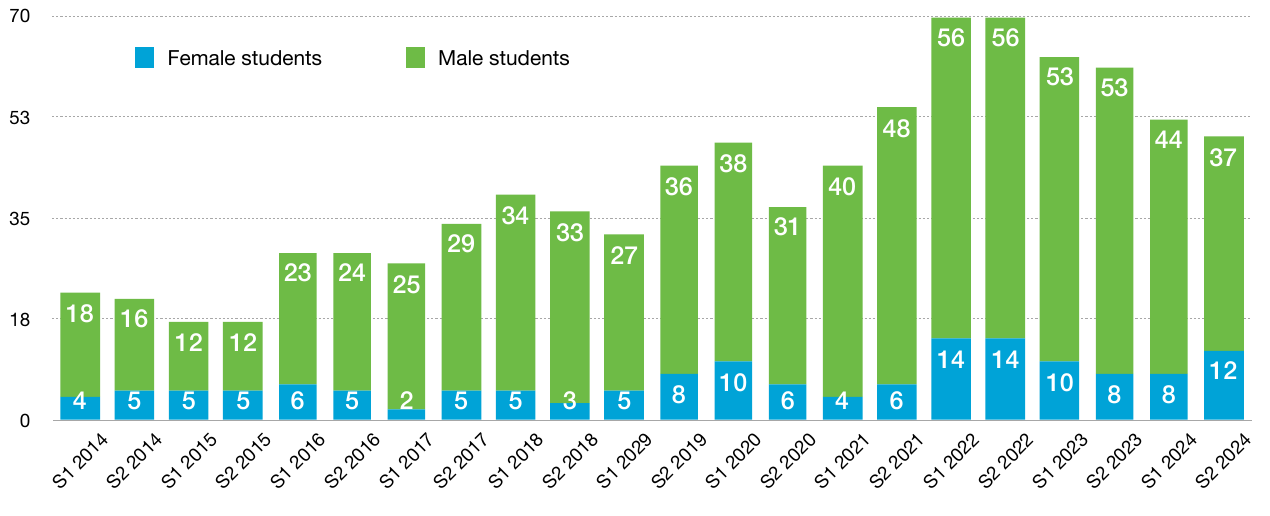}
   \caption{Student enrolments in the internship course over 2014-2024 (S1 and S2 denote Semester 1 and Semester 2 respectively).}
    \label{fig:internshipEnrolments}
\end{figure*}

\textbf{Eligibility Criteria.} To begin the internship, students need to have completed their first and second years in the Software Engineering program. The course relies on the foundational knowledge and understanding of processes, technologies, and applications acquired during those first two years.

\textbf{Preparation and Application Process.} 
Eligible students can search for internship positions through self-sourcing on platforms like Seek, LinkedIn, and Jora. However, applications must be approved by the course coordinator to ensure relevance to SE roles, full-time status, minimum wage compliance, and proper duration. Students can also consider jobs sourced by the Industry Engagement team, which have already been pre-approved by the course coordinator. Depending on the nature of the internship, students may need to submit their applications with necessary documents like their resume, cover letter, etc. Some companies, e.g., Leidos, NAB, WEX, Papercut, Astral, and Redcat, have long-term involvement in this activity and have provided internship placements on a regular basis, sometimes even leading to an ongoing position.

The course coordinator, in collaboration with the Career Support team, conducts workshops each semester to help students develop their resumes, write effective cover letters, and prepare for job interviews. 
As part of these workshops, the Industry Engagement team also organises guest speaker sessions featuring professionals from our industry partners who regularly hire interns. These speakers share valuable insights into their recruitment processes, expectations, and tips for success. This not only helps students better understand how to navigate job applications but also provides a valuable opportunity to expand their professional networks.

\textbf{Recruitment.} Since this course has a WIL component, successful candidates must first send their contact/offer letter to their course coordinator for approval. After approval, the WIL team can issue a third-party agreement (a so-called WIL agreement), between the university, the student and the employer for the entire duration of the internship. Once all parties sign the WIL agreement, students need to enroll in relevant courses, as they will be assessed and monitored during their internship by the course coordinator.

Some students begin their Bachelor's program mid-year (from S2) and can start their internship at that time. If a student has difficulty finding a placement, they can postpone their internship to start in their 6th semester. Consequently, each semester features two cohorts: (i) students starting their internship this semester and continuing into the next, and (ii) students finishing their internship from the previous semester.

\begin{table}[h!]
\caption{Student's placement information. Small denotes companies with fewer than 100 employees, Medium denotes companies with between 100 and 1,000 employees, and Large denotes companies with more than 1,000 employees.}
    \centering
        \fontsize{8pt}{8pt}\selectfont

    \begin{tabular}{|l|c|c|c|}
    \hline
         \textbf{Company Type} & \textbf{Small} &  \textbf{Medium} &  \textbf{Large}\\
         \hline
         Banking and Financial & 1 & 0 & \cellcolor{lightgray!50}14\\ 
         \hline
          IT Services and IT Consulting & \cellcolor{lightgray!50}10 & \cellcolor{lightgray!50}9 & 2\\
         \hline
          Manufacturing and Engineering & 2 & 3 & \cellcolor{lightgray!50}12\\
         \hline
          Medical and Pharmaceutical & \cellcolor{lightgray!50}4 & 1 & 0\\
         \hline
          Software Solutions and Development & \cellcolor{lightgray!50}6 & 6 & \cellcolor{lightgray!50}14\\
         \hline
          Telecommunications & 1 & 0 & 2\\
         \hline
         Others & \cellcolor{lightgray!50}5 & 0 & \cellcolor{lightgray!50}4\\
         \hline
         International & 3 & 0 & 0\\
         \hline
          
    \end{tabular}
    \label{tab:studentplacement}
\end{table}

Table~\ref{tab:studentplacement} summarises the placements of students in 2023 and 2024. The 99 students completed their internships at six major company types, totaling 49 companies, including 3 international ones. The “Other” category includes workplaces such as Business Supplies and Equipment, Newspaper Publishing, and Education, and the “International” category indicates the companies outside Australia. 
The cells highlighted in grey denote the company types covered by our data sample. 
Approx. \text{26\%} of the students completed their internships in Software Solutions \& Development companies, followed by IT Services \& IT Consulting companies with \text{21\%} and  Manufacturing \& Engineering companies with about \text{17\%}. Additionally, about \text{15\%} of the internships were done in the Banking and Financial sector companies. Companies in the rest of the sectors had \text{10\%} or fewer students completing their internships. Nearly half of the students (49\%) completed their internships at large companies, followed by 32\% at small companies and 19\% at medium-sized companies.
 
\textbf{Assessments.}
The internship course assessment comprises three main components: monthly reports, final live presentations, and feedback from the intern's manager (employer), designed to evaluate students’ professional growth, communication skills, and achievement of learning outcomes over the two-semester period.

Each student is required to submit regular reports as compulsory assessment tasks throughout the internship. The primary purpose of these reports is to enable the course coordinator to monitor students’ progress and overall experiences in their placements, allowing for timely intervention when necessary. Given that each semester spans four months, students are expected to submit one report per month, resulting in a total of eight reports across the two-semester internship, provided none are missed. These reports collectively account for 50\% of the total grade per semester. Students are expected to submit their reports within the designated semester timeline, as submissions outside this period are not required. Nonetheless, they remain under the supervision of the course coordinator and are encouraged to seek support or advice whenever needed.

In addition to the reports, students are required to deliver a live presentation at the end of each semester, which accounts for 30\% of the overall grade. This component is designed to strengthen students’ communication, time management, and presentation skills, as well as their confidence and ability to respond effectively to questions. During the presentations, students are expected to reflect on their internship experiences, including what is going well or not, any unexpected challenges, improvements in performance, newly acquired skills, a typical workday (tasks, meetings, or events), connections to course learning outcomes, and whether they view the placement as a potential future career path.

At the end of each semester, the intern's manager completes a feedback form evaluating each student’s strengths, weaknesses, areas for improvement, and overall progress. This evaluation accounts for 20\% of the total grade and plays a crucial role in shaping the overall learning experience and professional development of the students.

\subsection{Methodology for Students' Perceptions (RQ1 and RQ2)}
\label{sec:methodstd}

To answer RQ1 and RQ2, we conducted an empirical study to characterise the perceptions of SE students on a one-year internship program. In this section, we provide an overview of its methodological aspects, where the next two sections present the core findings of the study. 

The content of the courses in the Software Engineering program has been updated and refined every few years to incorporate the recent trends, as well as based on the feedback the course coordinators obtain from the industrial partners and from students. Therefore, we focus our analysis on the feedback from students regarding the current state of the art.

\subsubsection{Data Collection}\label{sec:datacollection}
Throughout the one-year internship spanning two semesters, students must submit 7–8 reports, 3 to 4 per semester, depending on the start and end dates of their placement.
The nature of these reports is close to surveys with open-ended questions, which allow us to obtain an in-depth view of students' progress and their perceptions. 
In these reports, students need to answer numerous questions related to their placement. However, this study only focuses on students' answers to the following questions:
\begin{itemize}[leftmargin=4ex]
    \item {\textit{Please tell us what course/s you found useful through your work (for this month), and explain briefly how you applied the learning from the course at your work.}}
    \item {\textit{What were your obstacles/challenges this month?}}

\end{itemize}
In 2023 and 2024, a total of 736 reports were submitted by 99 students, comprising 82 males and 17 females. Among these, 82 were domestic and 17 were international students, across eight rounds of report submissions.  
Our analysis addressing research questions RQ1 and RQ2 is based on 91 reports submitted by 13 students who consented to participate in the study, which represents approximately 12\% of the total report count. Table~\ref{tab:ParticipantsCohorts} provides a statistical overview for the 2023-2024 period. It includes the total number of student enrollments within each internship cohort, the number of study participants in those cohorts, and their percentage relative to the cohort size. Furthermore, a demographic breakdown of overall enrolments and consenting participants is presented by gender and residency status. Additionally, we present the total number of reports submitted during this period, the number of reports analysed based on participant consent, and the percentage of these reports relative to the total. As shown in Table~\ref{tab:ParticipantsCohorts}, study participants comprised 13\% of the 2023-2024 internship cohort, with a total of 13 participants out of 99 individuals.

\begin{table}[h!]
    \caption{Internship cohort coverage within the corresponding cohort.}
       \centering
        \fontsize{7.5pt}{7.5pt}\selectfont
    \begin{tabular}{l|ccc|c}  
    \hline
 & \multicolumn{3}{c|}{\textbf{Internship cohort}}\\
         &  Semester1 2023 &  Semester2 2023 &  Semester1 2024 & \\
         &   Semester2 2023&  Semester1 2024 &   Semester2 2024 & \textbf{Total}\\
         \hline
         Number of overall students’ enrollments &  47&  14&  38 & 99\\
         
         -- Number of Male|Female|Other &  39|8|0 &  13|1|0 & 30|8|0 & -\\
         
         -- Number of Domestic|Internationals &  40|7 &  11|3 & 31|7 & -\\

         Number of overall students’ reports &  351 & 99 &  286 & 736\\
         
         Number of study participants &  7&  1&  5 & 13\\

         -- Number of Male|Female|Other & 5|2|0 &  1|0|0 & 4|1|0 & -\\
         
         -- Number of Domestic|Internationals &  4|3 &  1|0 & 4|1 & -\\

         Number of study participants’ reports, & 46 & 8 &  37 & 91\\
         
         The percentage of study participants &  15\%&  7\%&  13\% & 13\%\\
         
         The percentage of study participants’ reports &  13\% & 8\% &  13\% & 12\%\\
        \hline         

    \end{tabular}
    \label{tab:ParticipantsCohorts}
\end{table}

\textbf{Demographic Breakdown of the Consenting Participants.} In this section, we present a demographic breakdown of the consenting participants, along with information about their roles, the tasks they were involved in, the type of company they were associated with, and the size of that company. All this information is summarised in Table \ref{tab:DemgparticipantBD} for clarity. As shown in the table, the study included 13 participants, comprising 3 females and 10 males. Among these participants, 4 are international students and 9 are local students. The majority of interns were involved in a wide range of technical and analytical activities spanning software development, testing, and data management. Their work commonly included software and system development, web and mobile application maintenance, and participation in agile workflows. Many contributed to quality assurance and automated testing, data analysis and reporting, and the development of dashboards and machine learning solutions. Others engaged in configuring devices, supporting software systems, and assisting with AI and data migration projects. Overall, the interns’ roles reflected a strong emphasis on practical software engineering, data-driven problem-solving, and collaborative development practices.

Our data sample is representative in terms of gender, status, and the type and size of internship companies. In our sample, the gender distribution includes 3 out of 13 participants who are female, which is approximately 23\%. This closely resembles the gender distribution of course enrolments for 2023-2024, where females make up 17 out of 99 students, or about 17\%. Additionally, the average for international students in the 2023-2024 internship course is 17 out of 99, approximately 17\%. This is similar to our data sample, which includes 4 out of 13 international students, making up 31\% of the sample. Out of our participants, 8 out of 13 (62\%) completed their internships at large companies. This is comparable to the overall statistics, where 48 out of 99 students (49\%) interned with large firms (see Table \ref{tab:studentplacement}). Our sample also reflects the types of companies where internships took place; 4 out of 13 participants (31\%) interned at Software Solutions and Development, which is close to the overall figure of 26\% for all students. Additionally, 2 out of 13 participants (15\%) worked in IT Services, IT Consulting, and Manufacturing and Engineering, which corresponds closely to the total number of students in these sectors (21\% and 17\% respectively).


\begin{table}[!ht] 
 \caption{Demographic overview of participants providing consent, along with their roles and responsibilities.}
 \label{tab:DemgparticipantBD}
 \centering
    \fontsize{8pt}{8pt}\selectfont
    \begin{tabular}{
        >{\raggedright\arraybackslash}p{0.2cm}  
        >{\raggedright\arraybackslash}p{1.4cm}  
        >{\raggedright\arraybackslash}p{1.2cm}  
        >{\raggedright\arraybackslash}p{3.5cm}  
        >{\raggedright\arraybackslash}p{2.2cm}  
        >{\raggedright\arraybackslash}p{3.8cm}  
    }
    \toprule
    \textbf{ID} & \textbf{Status} & \textbf{Gender} & \textbf{Company Type} & \textbf{Company Size} & \textbf{Role} \\ 
    
    \midrule
    
    \multirow{2}{*}{1} & International & Female & Software Solutions and Development & Small & Technical Business and Support Analyst Intern \\
     & \multicolumn{5}{p{10.5cm}}{\textbf{Responsibility:} Software testing and data analysis within an agile environment.} \\ 
     
    \midrule
    
    \multirow{2}{*}{2} & Domestic & Male & IT Services and IT Consulting & Small & Junior Web Developer Intern \\
     & \multicolumn{5}{p{10.5cm}}{\textbf{Responsibility:} Website and system development, maintenance, integration.} \\
     
    \midrule
    
    \multirow{2}{*}{3} & Domestic & Male & IT Services and IT Consulting & Medium & Developer Intern \\
     & \multicolumn{5}{p{10.5cm}}{\textbf{Responsibility:} React/React Native app development and maintenance.} \\
     
    \midrule
    
    \multirow{2}{*}{4} & Domestic & Male & Banking and Financial & Large & Technology Intern \\
     & \multicolumn{5}{p{10.5cm}}{\textbf{Responsibility:} QA, automated system testing, defect resolution.} \\ 
     
    \midrule
    
    \multirow{2}{*}{5} & International & Female & Software Solutions and Development  & Large & Product Engineering and Quality Intern \\
     & \multicolumn{5}{p{10.5cm}}{\textbf{Responsibility:} Software testing within an agile environment.} \\ 
     
    \midrule
    
    \multirow{2}{*}{6} & Domestic & Male & Medical and Pharmaceutical & Small & Software Engineer Intern \\
     & \multicolumn{5}{p{10.5cm}}{\textbf{Responsibility:} Development and maintenance of ML and web applications.} \\    
     
    \midrule
    
    \multirow{2}{*}{7} & International & Male & Software Solutions and Development  & Large & Digital Analytics Intern \\
     & \multicolumn{5}{p{10.5cm}}{\textbf{Responsibility:} Developing analytics dashboards, supporting data warehousing.} \\ 
     
    \midrule
    
    \multirow{2}{*}{8} & International & Male & Others  & Large & Junior AEM/Java Developer Intern \\
     & \multicolumn{5}{p{10.5cm}}{\textbf{Responsibility:} Developing, testing and maintaining AEM components.} \\ 
    
    \midrule
    
    \multirow{2}{*}{9} & Domestic & Male & Manufacturing and Engineering & Large & Logistics Continuous Improvement Intern \\
     & \multicolumn{5}{p{10.5cm}}{\textbf{Responsibility:} Generating daily reports, maintaining analytical tools \& scripts.} \\ 
     
    \midrule
    
    \multirow{2}{*}{10} & Domestic & Male & Software Solutions and Development  & Large & Data analyst Intern \\
     & \multicolumn{5}{p{10.5cm}}{\textbf{Responsibility:} Optimising SQL, analysing data, developing demos in Python.} \\ 
     
    \midrule
    
    \multirow{2}{*}{11} & Domestic & Female & Others  & Small & Junior Test Analyst Intern \\
     & \multicolumn{5}{p{10.5cm}}{\textbf{Responsibility:} Test case automation, functional testing, defect management.} \\ 

    \midrule
    
    \multirow{2}{*}{12} & Domestic & Male & Manufacturing and Engineering & Large & Software Innovations Intern \\
     & \multicolumn{5}{p{10.5cm}}{\textbf{Responsibility:} Device configuration, software support, development using C\#.} \\ 

    \midrule
    
    \multirow{2}{*}{13} & Domestic & Male & Software Solutions and Development  & Large & Junior Data Analytics Intern \\
     & \multicolumn{5}{p{10.5cm}}{\textbf{Responsibility:} Developing AI \& data analytics solutions, data migration tasks.} \\ 

    \bottomrule
    \end{tabular}
\end{table}

\subsubsection{Data Analysis}\label{sec:dataanalysis}
In this section, we describe our analysis process for RQ1 and RQ2.

\textbf{Quantitative Analysis for RQ1:}\label{sec:quananalysis}
We asked the students to disclose the courses they found valuable through their work every month. We evaluated the students' responses to this question to address \textbf{RQ1}. As the data collected from the students was in text format (approximately 7000 words), the first step was to extract the important features from it. Then, to bring consistency in the answers, we needed to standardise the responses. For instance, while answering the question about the courses they found valuable through their work, some students wrote incomplete course names, others used abbreviations, and so on. Therefore, we noted down all the relevant keywords and created a standard format to record the courses, which allowed us to count them and perform an analysis. The courses were also classified according to their content, the knowledge they provide, and the skills they aim to develop. We then shared all courses and their respective categories with the SE Program Manager for feedback. Based on the manager's input, we further refined the categories. Table~\ref{tab:coursecategories} presents the set of courses the students mentioned and the various categories to which courses can be assigned. For instance, \textbf{SE} denotes Software Engineering, while \textbf{Alg/Comp} denotes Algorithm and Computing, and \textbf{DS/AI} denotes Data Science and Artificial Intelligence. Courses related to programming are classified as \textbf{GP}, which stands for General Programming. On the other hand, programming courses that focus on web development, such as Full Stack Development and Programming Studio 1, are classified as \textbf{WP} (Web Programming). Additional information about course dependencies, prerequisites, the year of study in which each course is offered, and whether the course is classified as elective or not is presented in Table~\ref{tab:coursecategories}. Information about the content of all courses is included in our replication package \cite{replication} to support further studies.

\begin{table}[h!]
    \caption{Course category taxonomy. SE: Software Engineering, Alg/Comp: Algorithm and Computing, GP: General Programming, DB: Database, WP: Web Programming, DS/AI: Data Science and Artificial Intelligence, Y: Year, S: Semester, /: Or.}
    \centering
    \fontsize{8pt}{8pt}\selectfont
    \renewcommand{\arraystretch}{1.3}
    \begin{tabular}{l c c c}
    \hline
        \textbf{Course} & \textbf{Category} & \textbf{Prerequisite} & \textbf{Year}\\
         \hline
         Advanced Programming Techniques (APT) & GP & None & Y2-S1\\
         Programming Bootcamp 1 (PB1) & GP & None & Y1-S1\\
         Programming Bootcamp 2 (PB2) & GP & PB1 & Y1-S2\\
         Programming Studio 2 (PS2) & GP & PB2 & Y1-S2\\
         Software Engineering Fundamentals (SEF) & SE & PS1 & Y2-S1\\
         Software Engineering: Process and Tools (SEPT) & SE & SEF \& PB2 & Y2-S2\\
         User-centred Design (UCD)& SE & None & Y1-S1\\
         Full Stack Development (FSD) & SE, WP, DB & PS1/WP & Y2-S1\\
         Programming Studio 1 (PS1) & SE, WP, DB & PB1 & Y1-S1\\
         Web Programming (WP) & WP & PB1 & Elective\\
         Algorithms and Analysis (AA) & Alg/Comp & PB2/APT & Y1-S1\\
         Mathematics for Computing 1 (MC1) & Alg/Comp & None & Y1-S1\\
         Operating Systems Principles (OSP) & Systems & APT/PS2 & Y2-S2\\
         Data Communication \& Net-Centric Computing (DCNC) & Systems & None & Elective\\
         Cloud Computing (CC) & Cloud & APT/PS1 & Elective\\
         Introduction to Analytics (IDA) & DS/AI & None & Y1-S2\\
          Machine Learning (ML) & DS/AI & MC1/MC2 \& PS1 & Elective\\
         Mathematics for Computing 2 (MC2) & DS/AI & MC1 & Y1-S2\\
         Practical Data Science (DSc) & DS/AI & None & Elective\\
         Advanced Programming for Data Science (APDSc) & DS/AI & PB1 & Elective\\
         Business Data Management (BDM) & DS/AI & None & Elective\\
         \hline
    \end{tabular}
    \label{tab:coursecategories}
\end{table}

\textbf{Qualitative Analysis for RQ2:}
\label{sec:qualanalysis}
We asked the students to share the challenges they experienced during the internship in eight reports. We analysed the students' responses to this question using the thematic analysis technique \cite{braun2006using}, to answer \textbf{RQ2}. The combined responses from the students comprised approximately 5,000 words. Figure \ref{fig:codingprocess} shows sample responses provided by two students in their submitted reports. We also used the NVivo software to facilitate our qualitative analysis process. We first imported all students' responses (data) to NVivo. In the first step, the second author (coder) read the data to become familiar with the data. This step resulted in capturing key points in the students' responses. The next step focused on coding the data, which led to the construction of the initial codes. Figure \ref{fig:codingprocess} illustrates the examples of the key points extracted from two student responses, which then led to the identification of three codes. In the next step, the initially constructed codes were examined and grouped into potential themes. We next assessed the quality of the identified themes. In this step, we developed arguments to justify each of the extracted themes. Specifically, we examined whether the themes were sufficiently supported by the data, whether the supporting data demonstrated coherence or diversity, whether the boundaries between themes overlapped or reflected nuanced distinctions, and whether the themes captured the most important aspects of the data \cite{braun2021thematic}. In the final stage, each theme was assigned a clear title.

Once all \textit{codes} and \textit{themes} were identified, the first and third authors reviewed them. Next, several meetings between the coder and the first and third authors were organised to calibrate the \textit{codes} and \textit{themes}. Any disagreements and conflicts were resolved using the negotiated agreement approach as one of the widely used and highly cited methods in qualitative research to solve disagreements among coders \cite{campbell2013coding}.

\begin{figure*}
    \centering
    \includegraphics[width= 1\linewidth]{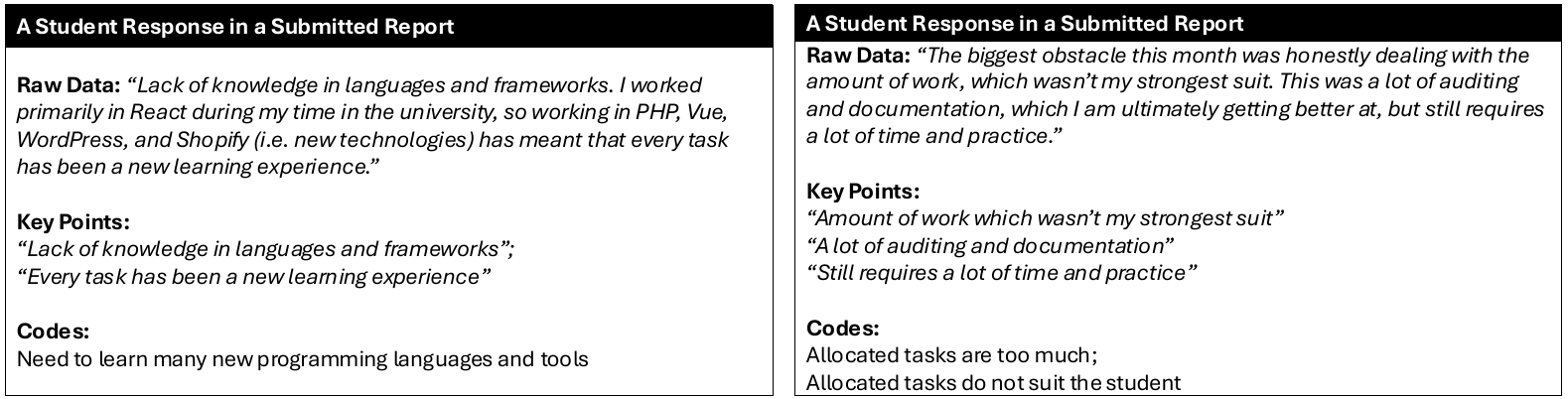}
   \caption{Examples of identifying key points and assigning codes to two student responses}
    \label{fig:codingprocess}
\end{figure*}


\subsection{Methodology for Educators' Lessons' Learned (RQ3)}
\label{sec:methodintershipredesign}
To address RQ3, this study draws on the experiences of the last two course coordinators, employing practice-based inquiry and iterative evaluation of the internship course. 
This method allows for a systematic examination of how the course has evolved over time and how specific instructional practices influence students' progress, the opportunities for the course coordinator to support the students, as well as the satisfaction of students and industry partners.
It incorporates insights gained from ongoing observation, student feedback, and institutional documents.

\subsubsection{Data Collection}\label{sec:datacollectionQ3}
The data that inform this methodology include reflections from coordinators, internship reports submitted by students, online forum discussions within the university’s e-learning system, and ongoing feedback from industry partners collected through Microsoft Forms over the past 10 years. Together, these sources have guided the continuous adaptation of the internship framework to address the evolving demands of the industry and to enhance the overall student experience.

It is worth noting that initially, internship activities were guided solely by tasks assigned by the host companies, with the university’s only requirement being a semesterly internship report. These 10–15 page reports, submitted using a Word template, asked students to reflect on their work duties, the application of academic knowledge, workplace conditions (e.g., readiness and training quality), challenges faced, tools and technologies used, skill development, and overall motivation. To support peer exchange, an online discussion forum was also created within the university’s e-learning system.

\subsubsection{Data Analysis}\label{sec:dataanalysis}
\label{sec:qualanalysisR3}
In the first step, the first and third authors independently reflected on their experiences delivering the internship course during their time as course coordinators, using the collected data as a foundation. They then held several meetings to share insights and discuss the course’s evolution, including the changes implemented and lessons learned over the years. Together, they analysed how the course structure had developed, with a focus on the goals and outcomes of each adjustment (see Figure \ref{fig:InternshipMidifications}). After summarising the key insights, a follow-up meeting with the second author was held to review the findings and refine the final lessons learned.

\begin{figure*}
    \centering
    \includegraphics[width= 1\linewidth]{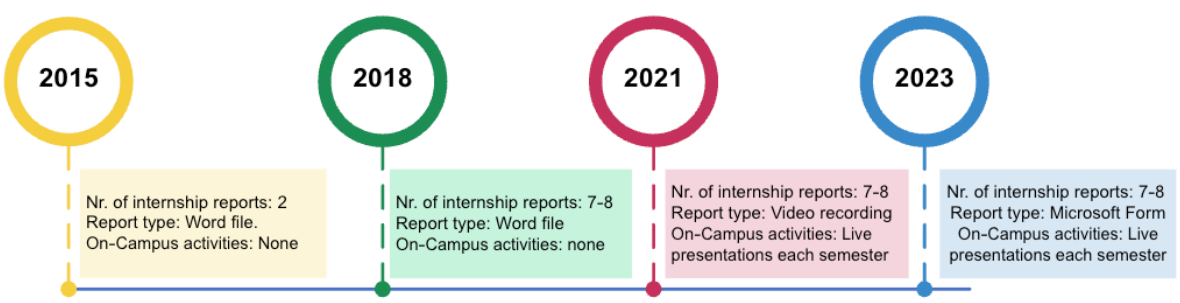}
   \caption{Adjustments of the internship settings over 2015-2025}
    \label{fig:InternshipMidifications}
\end{figure*}

\subsection{Limitations}
\label{sec:limitation}
This section summarises limitations that may have arisen in our research and their respective mitigation strategies.

Our research method in Section \ref{sec:quananalysis} involves manual steps to identify and categorise related courses and challenges, as detailed in Sections \ref{sec:findingsRQ1} and \ref{sec:findingsRQ2}, respectively. To address potential biases in the first one, all three authors, who possess extensive knowledge of SE courses, actively participated in the categorisation. Additionally, we presented all courses along with their corresponding categories to the SE Program Manager for feedback. Finally, we refined the categories based on the Program Manager's input. 
Furthermore, to mitigate potential bias in identifying challenges from the students’ reports, one author initially proposed three distinct challenge themes. The remaining two authors then thoroughly reviewed these challenge themes, engaging in multiple discussion sessions to resolve any disagreements. This collaborative review process ensured a comprehensive understanding of the challenges and improved the accuracy of their identification.

Another potential limitation could arise from the types of questions we asked the students to answer in their reports. These questions may not accurately reflect the courses they found relevant or the challenges they faced during their internships. Additionally, the students' reports were both mandatory and self-reported, which may have influenced our findings. Therefore, the conclusions drawn from this data may not be applicable to situations where participation is voluntary. Furthermore, the students who consented to have their reports used might have had a stronger interest in the internship program, potentially introducing selection bias. This suggests that the students' perceptions may not completely reflect the effectiveness of the internship program.

Similarly to the case of course evaluation surveys that are subject to gender and
race biases~\cite{mitchell2018gender,gordon2021role,khokhlova2023evaluating},  
students' perception of the course's contributions and importance for their preparedness to the internship, might be impacted by the personality of the academics who taught the course, their teaching style, gender or race. In our study, we assume that the course content might still play the dominant role for students' perception, which is a limitation of our analysis. 
As the content of the courses is updated every few years, we limited our analysis to the last two years to have a snapshot of the current state of the art. However, as students typically complete the courses one to two years before their internship, and the data were collected in 2023-2024, the recent impact of the rise of GenAI on course content, assessments, and students' perceptions hasn't been covered by our study.

Moreover, the identified courses and three challenge themes stem from an analysis of 91 reports submitted by 13 students (out of 99 students, i.e., approx. 13\%), who gave us consent to use their report data. It is also important to note that all findings are based on data collected from 2023 and 2024 at a single university. 
It is essential to acknowledge that these challenge themes may not encompass all possible students' perceptions, as other students may report different courses or concerns; that is, the challenges presented in this study may not accurately represent the full spectrum of difficulties that interns could encounter throughout their internships. Thus, while these findings provide valuable insights, they should be viewed as a sample rather than a comprehensive overview. 

Last but not least, a limitation arises from the number and variety of companies involved in the study. While a total of 99 students have had experience working across 49 different companies, the findings were from the data collected specifically pertaining to the activities observed in only 13 of those companies, i.e., only approx. 27\% have been covered. This limitation raises questions about the extent to which the findings may be representative of the broader range of companies within the original sample of 49 or even other companies outside of that group. Therefore, generalising the results could be challenging, as the activities and contexts of the 13 companies might not accurately reflect those of all the companies in the larger population.

\section{Findings}
\label{sec:findings}
\subsection{RQ1: What SE program’s courses do students believe most contributed to preparing them for the internship program?}\label{sec:findingsRQ1}

An analysis of 91 monthly reports from 13 consenting students indicates that Software Engineering: Process and Tools (SEPT) is perceived as the most relevant course throughout their internships. Students also frequently mentioned Programming Studio 1 (PS1) and Software Engineering Fundamentals (SEF) as useful in supporting their workplace tasks. In contrast, courses such as User-Centred Design (UCD) and Algorithms and Analysis (AA) were less frequently applied during internships.

As shown in Table \ref{tab:coursecategories}, which groups courses based on similarities in content, software engineering (SE)–related courses had the strongest impact, with eight students reporting their usefulness on a monthly basis. This category includes SEF, SEPT, UCD, PS1, and Full Stack Development (FSD). Courses in general programming (GP), data science/artificial intelligence (DS/AI), web programming (WP), and databases (DB) followed in usefulness. Detailed descriptions of course content and further supporting analysis are provided in our replication package \cite{replication}.

Among SE courses, SEPT appears to contribute most significantly to students’ internship performance. SEPT, an advanced second-year course, introduces the end-to-end software engineering lifecycle, including requirements analysis, software architecture design, version control, automated testing, deployment practices, and collaboration in agile environments. Its strong alignment with professional workflows likely explains its high relevance during internships.

SEF, also an advanced course in the second year, develops foundational software engineering skills by introducing the software development life cycle, agile methodologies, UML modelling, object-oriented design, testing practices, and team collaboration. PS1, a first-year course, was also frequently recognised for its usefulness. Despite its introductory level, PS1 gives students their first exposure to web technologies, version control, continuous integration, and database systems through collaborative studio-based projects—skills that directly transfer to industry contexts.

Overall, SEF, SEPT, and PS1 form the core foundation upon which students build their internship capabilities. While Year 1 courses focus on fundamental programming, Year 2 courses strengthen applied software engineering skills that better align with typical internship responsibilities. Students also frequently recognised courses in the GP and AI/DS categories as beneficial, given that many study participants were involved in coding and data analytics. Interestingly, most courses in the AI/DS category are electives, highlighting the growing demand for data-related skills in the industry.

We also acknowledge that the importance of some courses may not be fully recognised by students, as they tend to report only those that directly relate to their day-to-day internship tasks. For example, courses such as Algorithms and Analysis or other mathematics-based courses were mentioned less frequently; however, their underlying concepts remain essential. These foundational skills are embedded in activities such as programming, problem-solving, and analytical work, even if students do not explicitly attribute their relevance during the internship.

The curriculum structure appears to be well-aligned with market needs; students first build the essential foundations in programming, followed by advanced courses that support professional readiness. The results indicate that this progressive design effectively prepares students for the authentic software engineering tasks they encounter during internships.

\subsection{RQ2: What challenges do students experience during the
internship program?} \label{sec:findingsRQ2}
Our qualitative analysis of the student data identified three challenges (i.e., themes) that the students faced during the internship. To further support the identified challenges, each challenge will be described with a few representative quotations.
In what follows, we use a notation $\times$\emph{\textbf{i}} to denote that we identified \emph{\textbf{i}} responses reporting the corresponding challenge.

\textbf{Challenge 1:  \textit{Complexity of allocated tasks} ($\times$\textbf{48})}. We identified 48 reports indicating that students experienced difficulties with the nature of the tasks assigned to them. Some students pointed out that the recruiting company did not take into account their skills and expertise when allocating them to a task. They often complained that the complexity of the allocated tasks was higher than their current expertise and skills. Some students also complained about the volume of the (complex) tasks allocated to them. This led to some of them feeling pressure and not being able to cope well with the workload and maintain productivity. In the following quotation, the student indicates that the tasks assigned to him/her were complex, which made finding the correct solution challenging.
\begin{quoting}
\faQuoteLeft{} \textit{I found that one of the tasks assigned to me was too complicated, so I had to ensure that the output of the new refactored code matched the old code. I struggled to get the logic right to get the numbers to match.''} (Data Analyst Intern)
 \end{quoting}}
The following quotation illustrates that the number of assigned tasks was excessive, leading to a stressful situation for one of the students.
\begin{quoting}
  \faQuoteLeft{} \textit{My main obstacle was workload- there's been quite a lot to do this month, which has been stressful at times.''} (Junior Web Developer Intern) 
 \end{quoting}}

A common task that the students were asked to do, particularly at the beginning of their internship, was to \textbf{install and configure tools and infrastructures} they were required to use or rely on. Our observation shows that this was not a straightforward task for many students.
\begin{quoting}
\faQuoteLeft{} \textit{My major obstacles this month were setting up and installing the app on different printing devices for device testing of a new version release build.''} (Product Engineering and Quality Intern) 
\end{quoting}
\begin{quoting}
\faQuoteLeft{} \textit{Setting up the monitoring software was the biggest obstacle - many problems arose (mostly the issues with working with multiple OS + old/outdated frameworks and software), so I spent a lot of time debugging OS-specific issues.''} (Junior Web Developer Intern) 
\end{quoting}

Our analysis also shows that \textbf{getting access permission to use and/or configure infrastructure and tools} was another frequent challenge. 
\begin{quoting}
\faQuoteLeft{} \textit{One of the main issues, however, was performing jobs, which for me was configuring the devices and setting them all up on the same network for the customer, all of which required communication with the client's IT/security team due to complications with connecting to their network. Either it was their policies or firewalls, it made it difficult to access their network and make sure our devices were configured correctly as well as functioning properly.''} (Junior Data Analytics Intern)
\end{quoting}

Another complex task that students frequently discussed was \textbf{navigating large codebases} and \textbf{identifying the relevant parts} for their tasks. For example, a Quality Assurance Intern noted: \textit{“Specifically, [my problem was] learning the code base of my team's repository.'' The students also reported challenges, including adhering to the coding style of the codebase, refactoring existing code, and integrating their own code into the codebase. The two quotations provided below vividly illustrate these challenges. 
 \begin{quoting}
  \faQuoteLeft{} \textit{Things I had initially implemented worked; however, they did not integrate with the existing code base very well because of inconsistent APIs.''} (Developer Intern)
\end{quoting}
\begin{quoting} 
\faQuoteLeft{} \textit{I am finding migrating code to be an aspect that I find difficult. This is a task that I haven't done much of in the past, certainly not as part of university, and I find it hard to break down this task into larger sections.''} (Software Engineer Intern)
\end{quoting} }
%
Another task that was allocated to the students, which presented difficulties for them, was \textbf{testing and debugging} applications. These difficulties varied, ranging from not understanding how to debug massive applications to identifying root causes of errors in automation testing, replicating bugs, and increasing test coverage.
\begin{quoting}
   \faQuoteLeft{} \textit{This month, the document templates I was supposed to be testing are run through an Asynchronous service instead of just a synchronous service I am used to. So, it took a bit of time for me to understand how they needed to be tested in a different format.''} (Software Engineer Intern)
\end{quoting}
\begin{quoting}
 \faQuoteLeft{} \textit{I was presented with many browser-level compatibility issues, which were hard to debug due to a lack of similar situations online.'' (Developer Intern)}\end{quoting}

\textbf{Challenge 2:  \textit{Learn technologies and tools}} ($\times$\textbf{20}).
Another common challenge reported by the students was learning technologies (e.g., programming languages, testing frameworks) and tools required to do their allocated tasks. The students pointed out that they had little to no experience with some of the required technologies and tools, and spent significant time learning them. 
\begin{quoting}
\faQuoteLeft{} \textit{Lack of knowledge in languages and frameworks. I worked primarily in React during my time at university, so working in PHP, Vue, WordPress, and Shopify (i.e., new technologies) has meant that every task has been a new learning experience.''} (Junior Web Developer Intern) 
\end{quoting}
\begin{quoting}
\faQuoteLeft{} \textit{Google Analytics is a bit complicated and hard to understand. It's new for me, so I'm struggling to understand it.''} (Technical Business and Support Analyst Intern)
\end{quoting}

In some cases, the technologies and tools the students had to use were either old, custom, or developed by the company without proper documentation, making the learning process more difficult and time-consuming. 
\begin{quoting}
   \faQuoteLeft{} \textit{WPF and the custom library that gets the IND700 code working will be the only two areas that are 'new'; that is, they will be more of an unfamiliar concept than the others. The custom library, named Proworks, is the most outlier area. The two seniors of my team had to travel to Germany in order to get day 1 training for this concept, so learning it will be hard.''} (Junior Data Analytics Intern)
\end{quoting}
\begin{quoting}
   \faQuoteLeft{} \textit{I also found great difficulty in learning the underlying custom systems and frameworks the company had developed for itself.''} (Junior Web Developer Intern)
   \end{quoting}

\textbf{Challenge 3:  \textit{Teamwork and Support Barriers}} ($\times$\textbf{20}). 
The students attributed some of the obstacles they experienced to their poor communication and collaboration with the rest of their team members. Communication with offshore team members was also mentioned by some students as a challenge, especially due to the time zone difference. 
 \begin{quoting}
  \faQuoteLeft{} \textit{Since the team zone delay is -9 hours, communication with knowledgeable colleagues is difficult.''} (Software Engineer Intern)
\end{quoting}

We also observed that students sometimes found it difficult to communicate and collaborate with clients and other stakeholders.  
\begin{quoting}
   \faQuoteLeft{} \textit{To implement the digital data layer, we need developers to help implement JavaScript in the backend of web code. So, explaining to them how the data layer works and what they need to do at the backend is quite complex.''} (Digital Analytics  Intern)
\end{quoting}

Our data includes some references showing that the students sometimes did not draw enough support from the recruiting company and their (senior) team members when they needed help. Some students also perceived that they were not adequately trained during their internship. 
\begin{quoting} 
\faQuoteLeft{} \textit{ProdX, LabX and FWN are software that were handled by a colleague in a different-ish department from my current team and had gone on long-service leave for 1-2 months, during which I was handling all enquiries and jobs that were delegated to him.''} (Junior Data Analytics Intern)
\end{quoting} 
\begin{quoting} 
\faQuoteLeft{} \textit{I needed to learn most of the skills myself, and with the help of online resources about client data layer implementation.''} (Data Analytics Intern)
\end{quoting}

Some students referred to the lack of support by referencing the documentation that was not available for completing their allocated task. They argued that some projects had insufficient documentation required to understand their functionality in-depth, posing difficulties for them to do their assigned tasks. In some cases, the students complained that the required documentation was entirely missing.
\begin{quoting}
 \faQuoteLeft{} \textit{I had to reverse-engineer an API and understand how it was meant to be used, due to lack of documentation.''} (Web Developer Intern)
\end{quoting}

\begin{quoting}
 \faQuoteLeft{} \textit{Only the seniors were able to travel and be trained properly for the new devices, so the rest of the team will have to make do with documentation and their input. This is a double-edged blade, as while communication between team members is good, since we are not experienced yet, anything any of us do not know, we will struggle with. This usually points to small intricacies that the manuals and documentation fail to explain, or when issues not covered pop up as well.'' } (Software Engineer Intern)
\end{quoting}


\subsection{RQ3: What lessons have been learned by the course
coordinator team over the last 10 years?}\label{sec:findingsRQ3}

\textbf{Lesson Learned 1:} \textbf{\textit{Students require live-contact (in-class or at least real-time online) with each other to increase their feeling of belonging to the university community and to allow observing how their peers are performing and what their progress is. }} 
An online forum doesn't effectively address this issue. Course coordinators also noted that, \textit{``These interns are still full-time students completing a one-year internship in the workplace, and they may feel that they are no longer students and do not belong to the university, which is not ideal.''}. We propose a solution that has been well-received by students: in addition to submitting reports, each semester students must deliver a live presentation, totalling two throughout the internship. This helps improve their communication skills, time management, slide content, and readiness for questions. During presentations, students discuss their work experiences, including successes and challenges, skills learned, typical duties, learning outcomes, and future career considerations. This activity fosters collaboration and networking among students, allowing them to share solutions to common issues. Presentations also expose students to various roles they might not have known about, prompting discussions and possibly influencing career choices. For instance, some students were inspired to switch employers after learning about appealing positions during these presentations. Prior to 2021, when live presentations were not part of the course, some students expressed, \textit{``We want to know what other interns are doing and how they handle the challenges.''} After live presentations were introduced in 2021, many students commented, \textit{``It is very interesting to hear about other interns' responsibilities and tasks; I want to learn more about this area because it is very interesting''}.

\textbf{Lesson Learned 2:} \textbf{\textit{Regular (e.g., monthly) reports are beneficial for students' progress. Reports should be textual, collected using an online form.}}
While receiving reports semester-wise (see Figure \ref{fig:InternshipMidifications}) allowed to have a reasonable overview of students' placements, it didn't allow the course coordinators to get timely feedback on students' perceptions of their progress and react promptly if students might need additional support. An intern from the 2015-2017 batch, where interns were only required to submit one report at the end of each semester, expressed, \textit{``I wish I had more regular contact with my course coordinator to discuss my concerns earlier in the semester and feel more supported instead of being on my own''}. Also, without being prompted by a course coordinator, students might hesitate to discuss their concerns or decide to postpone the discussion. To solve this issue, the students (1) should be contacted by the course coordinator on at least a monthly basis or (2) have to submit their reports more regularly, e.g., monthly. The first option has a number of critical disadvantages: 
\begin{itemize}[leftmargin=4ex]
    \item Students might perceive being micromanaged, 
    \item It might be hard for some students to express their concerns within a catch-up discussion,
    \item  This could significantly increase the coordinator's workload, especially with cohorts of over 50 students.  
\end{itemize} 
Therefore, our solution was to proceed with  {monthly reports}: each student usually submits 7-8 reports over their internship period, depending on the start times of their placement. These reports enable the course coordinator to monitor students’ progress and overall placement experience, ensuring they receive appropriate mentoring, are not overwhelmed by their duties, and that their assigned tasks align with the intended scope of the internship. Furthermore, Interns should reflect on their learning, identify challenges or skill gaps, and consider relevant courses to support their professional development during their final year of the program. Reports should discuss the students’ work duties and how they applied their academic knowledge in practice, as well as evaluate workplace factors, including readiness, communication, guidance, task distribution, and quality of training. Students are further expected to describe any obstacles they encountered, the technologies and tools they used, how their skills have improved, and their level of motivation and engagement within the role.

Over a number of semesters, we trialled video reporting: each student had to submit a 5-minute-long video recording of themselves, reporting their internship experience. While this solution might also be a good option to proceed with a smaller cohort, we decided to switch back to written reports, as this allows for better analysis of larger cohorts, including a comparison of students' progress and their perception of the placement situation. According to the course coordinator during 2021, \textit{``Although reviewing students’ progress and reflections through video recordings is highly valuable, it requires significant time to download and review them.''}. One intern from this cohort noted, \textit{``Recording even a 5-minute video is very challenging for me and takes a considerable amount of time.''}
Thus, based on our experience, the following approach is especially effective: 
Each student submitted their report using Microsoft Forms (or equivalent tool) to be reviewed by the course coordinator. 

\textbf{Lesson Learned 3:} 
\textbf{\textit{The increase in competitiveness levels is due to the decline in the number of interns recruited by big companies.}} As could be seen from Table \ref{tab:ParticipantsCohorts}, the number of interns has decreased over the last two years. This is the result of the general situation in the SE industry sector - the number of students recruited by companies has decreased. 
This reduction is primarily due to large firms, particularly in the banking and finance sectors, choosing not to recruit any interns at all. Instead, they have shifted their focus to graduate programs, which provide opportunities for recently graduated students rather than for SE students seeking internships in the middle of their studies. One of our industry partners noted that \textit{``employers often prefer interns who can transition into permanent positions instead of returning to university to complete their degrees''.} Another industry partner pointed out, \textit{``We invest in training them, but then they have to go back to university since they are still full-time students''.} Consequently, this trend has led to increased competition among students eligible to apply for internships at mid-sized or smaller companies. 

A decrease in the number of available positions was observed during the COVID-19 pandemic, prompting to seek solutions to address the challenging job market. Additionally, as university training shifted predominantly online, many international students were compelled to return to their home countries and attend classes remotely. A discussion between the course coordinator and program manager was \textit{``Everyone has been overwhelmed by the pandemic, and it is better to give international students the opportunity to return to their home countries, where they may have better internship opportunities and family support''}. Therefore, it has been decided to provide international students with the opportunity to secure internship positions in their home countries, contingent upon approval from the course coordinator and meeting other relevant degree requirements. Drawing from the experiences during the COVID-19 pandemic and the recent decline in the job market, we have now broadened this opportunity to all students, allowing them to seek internship positions abroad.

\section{Recommendations}
\label{sec:recommendations}

In this section, we integrate the study’s findings with the existing literature and present implications and recommendations for the two key stakeholder groups: \textbf{\textit{universities}} and \textbf{\textit{industry.} } Figure~\ref{fig:recommendations} presents an overview of these recommendations.  


\begin{figure*}
    \centering
    \includegraphics[width= 0.8\linewidth]{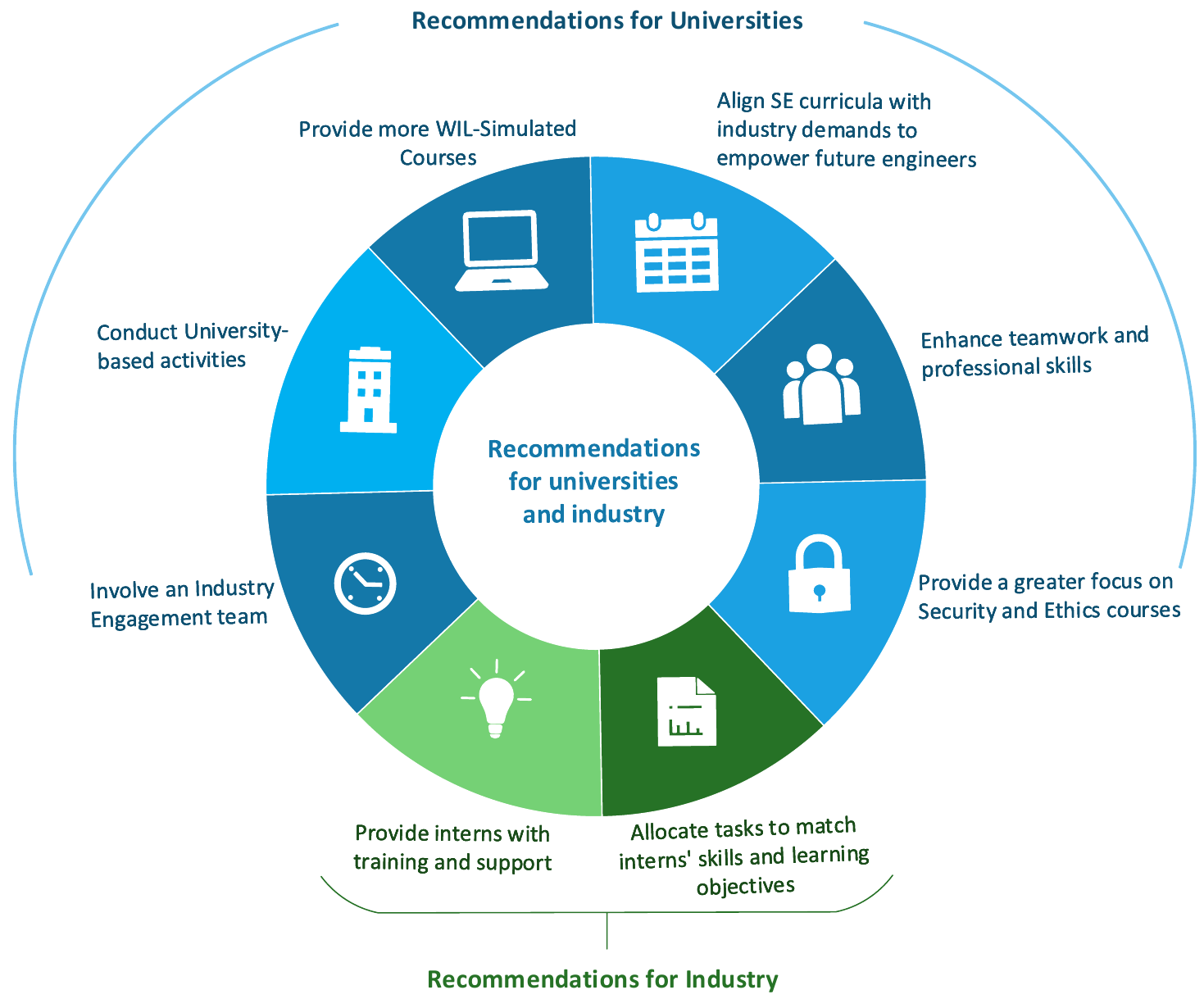}
   \caption{Overview of the proposed recommendations.}
    \label{fig:recommendations}
\end{figure*}

\subsection{Recommendations for Universities}

Drawing on Vygotsky’s Sociocultural Theory of Learning \cite{vygotsky1978mind}, which emphasises scaffolding and social interaction as essential for developing higher-order thinking, it becomes evident that internship experiences must be intentionally designed to provide guided participation and collaborative problem-solving. Kolb’s Experiential Learning Theory \cite{kolb2005learning} further reinforces the value of iterative, hands-on learning cycles that allow students to engage with real technologies and complex systems experiences that are crucial for professional readiness but often limited in classroom settings. This interpretation aligns with Reich et al. \cite{reich2015engineers}, who argue from a practice theory perspective that engineers’ professional learning emerges primarily through participation in authentic workplace practices rather than through formal instruction.

To connect these theoretical perspectives to our findings, it is important to recognise that Challenges 1, 2, and 3 reflect not only individual student difficulties but broader structural gaps in how practice-based learning is currently designed. Sociocultural and experiential learning theories help explain why these challenges persist and highlight the pedagogical conditions necessary for effective workplace preparation. This provides a foundation for understanding how internship experiences could be intentionally structured to support students’ professional growth.

With this theoretical grounding established, we now turn to how professional readiness in software engineering can be enhanced through deliberately designed internship experiences. Our findings suggest that students require structured and supportive environments that allow them to engage meaningfully with industry-relevant tools, practices, and collaborative processes. In particular, Challenges 1 and 3 point to the importance of exposing students to the scale, complexity, and real consequences of software engineering decisions elements that are difficult to reproduce in traditional coursework but are central to developing long-term professional competence. Building on these insights, this section outlines six recommendations aimed at \textbf{\textit{strengthening professional readiness}} as follows:

\faLightbulbO\ ~ \textbf{\textit{Provide more~WIL-Simulated Courses.}} One effective way for educators to enhance course efficiency and align with market demands is to integrate more WIL-simulated courses into their curriculum, a strategy endorsed by other researchers \cite{anivcic2022teaching,jackson2021embedding,spichkova2019industry,patacsil2017exploring}. These courses are designed to replicate real-world professional environments and tasks within an academic setting. Their goal is to bridge the gap between theoretical knowledge and practical application by simulating workplace experiences for students. WIL courses incorporate industry-relevant content, emphasise collaboration and problem-solving, and provide mentorship and feedback from instructors or industry experts. By participating in these courses, students can develop essential skills, build confidence, and better prepare for their careers, addressing challenges reported in \emph{Challenges 1 and 2}. Although providing these types of courses requires significant effort from course coordinators, particularly in designing the curriculum and finding suitable industry partners who are also passionate about this involvement, it ultimately has a profound impact on students' learning and their success in securing internships and performing well during their internships.


\faLightbulbO\ ~ \textbf{\textit{Align SE curricula with industry demands to empower future engineers.}} 
In \emph{Challenge 2}, students often expressed a need to learn various technologies, such as programming languages and testing frameworks, due to their limited prior experience with these tools. As highlighted in Section \ref{sec:findingsRQ1}, and consistent with prior research \cite{anivcic2022teaching,minnes2021cs,kapoor2019understanding,patacsil2017exploring}, this points to a critical responsibility for universities offering software engineering programs to ensure their curricula remain responsive to the rapidly evolving demands of the job market. Therefore, universities should provide comprehensive and regularly updated training in core areas such as software engineering, programming languages, artificial intelligence (AI), data science, and database management. This continuous alignment with industry needs can be achieved through active consultation with professionals and companies in the field. Embedding such practices not only enhances curriculum relevance but also equips students with the practical knowledge and skills required to navigate and succeed in dynamic technological environments.

\faLightbulbO\ ~ \textbf{\textit{Enhance teamwork and professional skills.}} 
Our software engineering program does not include a dedicated course focused specifically on professional skills, relying instead on group-based assignments across the degree to support students’ development in these areas. Nevertheless, one of the challenges identified was students’ difficulty in communicating and collaborating effectively with their team members (\emph{Challenge 3}), suggesting that this indirect preparation may be insufficient to equip students to work cohesively in professional environments. Such capabilities are essential for workplace success and may require targeted preparation prior to the commencement of the internship. Moreover, many workplaces do not provide structured training in these areas during onboarding or afterwards. To address this gap, universities should provide students with dedicated and targeted learning and development in teamwork, communication, and professional conduct. This recommendation aligns with other studies \cite{chng2018redesigning, warrner2021integrating,johnson2015framework}, highlighting the importance of professional skills courses, such as collaboration, effective communication, and conflict resolution, in preparing individuals to contribute positively in a team setting. Educators should prioritise incorporating training in these areas into the curriculum to better prepare students for workplace dynamics. According to our industry partners, who regularly hire interns from our program, technical skills remain important, but many companies place even greater value on candidates who demonstrate strong teamwork capabilities during recruitment and interviews.

\faLightbulbO\ ~ \textbf{\textit{Provide a greater focus on the Security and Ethics courses.}} The findings in Section \ref{sec:findingsRQ1} indicate that, from the students’ perspective, security-related and ethics-focused courses currently play no meaningful role in preparing them for software engineering positions or internships. This gap may stem from the specialized nature of the roles students pursue or the structure and delivery of these courses within universities. To address this, educators should reconsider both the curriculum and pedagogical approaches to enhance practical relevance through real-world case studies, stronger alignment with industry practices, and a clearer emphasis on the importance of security and ethics in professional software development. According to \cite{smith2023incorporating}, structural barriers continue to hinder the integration of ethics into computing curricula, despite broad educator support. Strengthening institutional support mechanisms is therefore essential to align ethics education with professional expectations. Similarly, \cite{towhidi2023aligning} argue that Information Systems curricula require redesign to meet industry needs and better prepare graduates for careers in cybersecurity. By adopting these changes, such courses can more effectively equip students to navigate the complexities of modern software engineering while also fostering professional responsibility and ethical awareness.


\faLightbulbO\ ~ \textbf{\textit{Conduct University-based activities.}} 
Both \emph{Lessons Learned~1 and 2} focus on how university-based activities are organised within an internship course. These activities play a vital role in boosting students' confidence and helping them navigate the challenges they face. A live presentation, along with opportunities to network, allows students to discuss their achievements and issues not only with the course coordinator but also with their peers. This interaction can help students manage the complexity of their assigned tasks, learn new technologies more effectively, and sometimes even find a more suitable internship placement. 
Additionally, regular reports can enable students to communicate their support needs to the course coordinator. For instance, when a student encounters a significant challenge, early intervention can be provided, ensuring that they do not have to address internship-related problems on their own and that support is not delayed until the end of the semester.

\faLightbulbO\ ~ \textbf{\textit{Involve an Industry Engagement team.}} 
As discussed in \emph{Lesson Learned 3}, the main factor affecting internship availability has been the decline in the number of positions offered by major companies, leading to increased competition for the available roles. To address this challenge, enhancing staffing within the \textit{Industry Engagement team} would improve outreach efforts to source more internship opportunities and develop long-term partnerships with reliable industry organisations that can consistently offer placements each year. However, these strategies require ongoing investment. As outlined in the \textit{Preparation and Application Process} of Section~\ref{sec:intenship}, the internship placement process is resource-intensive. 
The process involves more than just the \textit{Industry Engagement team} sourcing internship positions. It includes multiple stages, such as conducting resume, cover letter and interview preparation workshops, arranging networking opportunities, collecting feedback on selection criteria from industry partners, and providing ongoing support until students secure placements. This support comes from various parties, including the \textit{course coordinator}, \textit{program manager}, and the \textit{Industry Engagement team}. Moreover, once placements are confirmed, the \textit{WIL team} prepares the necessary agreements, and the course coordinator oversees the students throughout their year-long internships. While some universities have faced financial pressures that have led to reduced or reallocated support for these initiatives, our institution remains committed to sustaining the staffing and infrastructure necessary for effective internship coordination and supervision. Therefore, the challenges we face primarily stem from external market conditions rather than a decrease in institutional investment.

\subsection{Recommendations for Industry}

\faLightbulbO\ ~ \textbf{\textit{Provide interns with training \& support to help them navigate challenges and turn obstacles into valuable learning experiences.}} 
Students frequently reported challenges related to the initial setup and operation of tools and infrastructure (\emph{Challenge 1}), struggling with the steep learning curve of new technologies (\emph{Challenge 2}), and feeling unsupported due to inadequate technical assistance from the recruiting company and senior team members during critical moments (\emph{Challenge 3}). These obstacles left many interns struggling to meet expectations while navigating unfamiliar systems. To address these concerns, it is highly recommended that companies provide comprehensive, hands-on training sessions tailored to the tools and processes interns will encounter, also endorsed by others \cite{hardie2018value}. These sessions should not only familiarise interns with technical requirements but also instil confidence in their ability to perform effectively. Furthermore, continuous mentorship programs and well-structured learning opportunities should be integrated into the internship experience, enabling interns to build their skills progressively and develop professionally. Implementing a structured support framework, including regular one-on-one check-ins and guidance from senior staff, would create a safety net for interns as they face and overcome challenges. By offering timely assistance and constructive feedback, companies can empower interns to transform setbacks into valuable learning experiences. These findings also align with \cite{zehr2020student}, which emphasizes that both students and supervisors benefit from formal preparation or training prior to the internship.

\faLightbulbO\ ~ \textbf{\textit{Allocate tasks to match interns' skills and learning objectives while ensuring a balanced workload.}} 
Another common challenge identified by students during their internships pertains to the nature of the tasks assigned (\emph{Challenge 1}). Many students have reported concerns that recruiting companies often fail to adequately consider their individual skills and areas of expertise when allocating responsibilities. This misalignment can diminish engagement and hinder the overall quality of the learning experience, as also reported by other researchers \cite{p2014building,hacker2016permitted,grant2018ubiquitous}. To address these challenges, it is essential for task assignments to be thoughtfully designed to align with interns’ existing competencies and personal learning objectives. Such an approach would not only promote equitable workload distribution but also ensure that each intern has the opportunity to undertake responsibilities conducive to their professional development. By adopting a more tailored task allocation strategy, companies can enhance both intern engagement and learning outcomes, fostering a more enriching internship experience.

\section{Conclusion and Future Work}
\label{sec:conclusions}

This paper discusses students' perceptions of SE internships and the insights gained from a decade of supervising them. 
We discuss questions on \textbf{how} to structure an internship course, \textbf{how} to embed internships into the study curriculum, and \textbf{what skill set} might help students to be more successful in their internship journey.
Our study highlights the challenges students face and analyses strategies to address them. The study examines which academic courses best prepare students for industry roles and offers recommendations for educators and companies, such as aligning the curriculum with industry needs and emphasising teamwork and ethics. Although the findings are limited to a single institution, future work could involve a multi-university study to yield broader insights. Collaboration with other universities and industry partners is encouraged to further enhance internship programs.

\section*{Acknowledgement(s)}

This article benefited from the use of Grammarly and GPT-4 for editing and language improvement, ensuring a clear and effective presentation and communication of ideas. 

\section*{Disclosure statement}

No potential conflict of interest was reported by the author(s).

\section*{Ethical guidelines}

Ethics approval for this study was granted by the College Human Ethics Advisory Network (CHEAN) at RMIT University, Melbourne, Australia (Project No. 27119).

\balance
\bibliographystyle{acm}
\bibliography{main_bib}

\end{document}